\documentclass[aps,prl,twocolumn,superscriptaddress]{revtex4}

\usepackage{graphicx}
\usepackage{dcolumn}
\usepackage{bm}
\usepackage{amsmath,amssymb,amsbsy}
\usepackage{url}

\newcommand{\derivep}[2]{ \frac{\partial #1}{\partial #2} }
\newcommand{\bfA}{{\mathbf{A}}}

\newcommand{\bfE}{{\mathbf{E}}}
\newcommand{\bfF}{{\mathbf{F}}}
\newcommand{\bfH}{{\mathbf{H}}}
\newcommand{\bfJ}{{\mathbf{J}}}
\newcommand{\bfe}{{\mathbf{e}}}
\newcommand{\bfp}{{\mathbf{p}}}
\newcommand{\bfr}{{\mathbf{r}}}
\newcommand{\bfv}{{\mathbf{v}}}

\newcommand{\rmd}{{\, \mathrm d}}
\newcommand{\rme}{{\, \mathrm e}}
\newcommand{\rmi }{{\mathrm i}}

\newcommand{\cV}{{\mathcal V}}

\newcommand{\NN}{{\mathbb N}}

\newcommand{\RR}{{\mathbb R}}

\newcommand{\ZZ}{{\mathbb Z}}

\newcommand{\rot}{{\mathrm{rot}} \, }

\begin{document}

\def\Thales{Thales Electron Devices, rue Lat\'eco\`ere, 2, FR-78140 V\'elizy, France}
\def\Marseille{UMR 7345 CNRS--Aix-Marseille-Universit\'e, 
campus Saint-J\'er\^ome, case 321, \\ 
av.\ esc.\  Normandie-Niemen, FR-13397 Marseille cedex 20}
\def\Saratov{Saratov State University, Saratov 410012, Russia}

\title{Hamiltonian description of self-consistent wave-particle dynamics in a periodic structure}

%

\author{Fr\'ed\'eric Andr\'e}\email[corresponding author : ]{frederic.andre@thalesgroup.com}
\affiliation{\Thales}
\author{Pierre Bernardi} 
\affiliation{\Thales}
\affiliation{\Marseille}
\author{Nikita M. Ryskin}
\affiliation{\Saratov}
\author{Fabrice Doveil}
\affiliation{\Marseille}
\author{Yves Elskens}
\affiliation{\Marseille}



\date{\today}

\begin{abstract}
  {Conservation of energy and momentum in the
    classical theory of radiating electrons has been a challenging
    problem since its inception. We propose a formulation of classical
    electrodynamics in Hamiltonian form that satisfies the Maxwell
    equations and the Lorentz force. The radiated field is represented
    with eigenfunctions using the Gel'fand $\beta$-transform. The
    electron Hamiltonian is the standard one coupling the particles
    with the propagating fields. The dynamics conserves energy and
    excludes self-acceleration. A complete Hamiltonian formulation
    results from adding electrostatic action-at-a-distance coupling
    between electrons.}
  \newline 
  PACS numbers: 84.40.Fe (microwave tubes) \newline %
  52.35.Fp (Plasma: electrostatic waves and oscillations) \newline %
  52.40.Mj (particle beam interaction in plasmas) \newline %
  52.20.Dq (particle orbits) \newline %
%
\noindent
Keywords : {wave-particle interaction, traveling wave tube, $\beta$-transform, Floquet boundary condition}

\end{abstract}


\maketitle

To model consistently the interaction between electrons and waves in
devices such as traveling wave tubes, free electron lasers or
synchrotrons, we are presently left with two options.  The first
\cite{Gilmour94,Kartikeyan04,Taflove95} is to consider the flow of
electrons as a distributed charge and current density coupled with the
field through the Maxwell equations. Since it generates diverging
singularities, the particle nature of electrons is intentionally
overlooked until the question of determining the trajectories of the
flow is raised. {For the latter,} the only possibility is to
return to a particle description in which the Lorentz force
applies. This change of model for the flow precludes the description
of the wave-electron system in Hamiltonian form. One reason is that a
procedure is needed to distribute the electron charge and current into
a finite volume.  This procedure, usually based on meshing space, can
only be arbitrary.  The second option
\cite{Hartemann02,Jackson99,LandauLifshitzVol2} is to consistently
consider electrons as particles and to determine the field they
radiate starting from the Li\'enard-Wiechert potentials.
{Dirac's \cite{Dirac38} sharp analysis indeed provides an
  accurate determination of the reaction from the radiated field in
  the limit of a point electron, yet this approach leads to serious
  difficulties
  \cite{Dirac38,Jackson99,LandauLifshitzVol2,Spohn04,Thirring03};
  among these an infinite rest mass of the electron, self-acceleration
  and acausality, also appear incompatible with the existence of a
  well-posed Hamiltonian.}

Hamiltonians are essential to consistently define energy and momentum.
They also have great practical usefulness to find approximate
solutions in complex systems or to control errors in numerical
integration schemes \cite{Hairer02}. So their absence in the case of
the classical wave-electron interaction is both theoretically and
practically unsatisfactory. While the final solution to these problems
may involve (upgraded, regular) quantum electrodynamics, one could at
least {hope for} a consistent classical approximation, which would be the
classical limit of its quantum counterpart. For instance, in the limit
where the electron radiates a large number of photons, each of them
having a small energy compared with its kinetic energy, which is the
case for the aforementioned devices, one could expect that averaging
the quantum theory over a large number of emitted photons would
provide a consistent classical limit. But such a theory, free of the
above problems, remains to be found
\cite{Heitler36,Spohn04,remarque,Tiggelen12}. 

To progress in this direction, we
undertake the construction of a Hamiltonian {in situations
  where the boundary conditions are periodic in one or more
  dimensions, and we focus on the radiated field modes interacting with the electrons. 
  Our method applies both to traveling wave tubes, 
  meant to amplify ``slow'' modes in the periodic structure, 
  and to free electron lasers where the electrons couple to ``fast'' modes. }


First of all, one can distinguish the space-charge field, which is the
curl-free part of the fields, from the radiated field, which is the
divergence-free part. In free space, the space-charge field between
several electrons with positions $\bfr_i$ derives from the Hamiltonian
$\mathcal{H}_{\mathrm{sc}} 
= \frac{1}{8 \pi \epsilon_0} \sum_{i \neq j} e^2 / \| \bfr_i-\bfr_j \|$. 
Adopting this point of view
eliminates the problem of infinite rest mass, by  eliminating altogether
the electrostatic field degrees of freedom. So one is left with
finding a Hamiltonian for the radiated fields coupled with the
electrons. This is the object of this work.


The geometry is periodic along the $x$ axis with period $d$
{(extension to two- and three-dimensional lattices is
  straightforward)}.  The lattice may include perfect metal ($\bfE
\times \bfe_\perp = \mathbf{0}$, $\bfH \cdot \bfe_\perp = 0$) boundary
condition. All equations are written in the reference frame of these
boundary conditions. All periods, with shape $\cV_0$ and volume $|
\cV_0 |$, form a domain $\cV_\ZZ := \cup_{n \in \ZZ}\{\bfr + n d\bfe_x
: \bfr \in \cV_0\}$.  The propagating waves are described with their
electric and magnetic fields, $\bfE(\bfr,t)$, $\bfH(\bfr,t)$, obeying
the Maxwell equations with sources \cite{Bernardi11}.


We decompose the radiated field on the function basis made of the
propagating modes. These modes satisfy the metal boundary condition
{on cavity walls,}
and the Floquet condition, $\bfE(\bfr + d \bfe_x) = \rme^{-\rmi \beta
  d} \bfE(\bfr) $ for some $\beta \in \RR$.  Inside $\cV_0$, they
satisfy $\nabla \cdot \bfE_{s\beta}=0, \nabla \cdot \bfH_{s\beta}=0$,
and
\begin{eqnarray}
  \rot \bfE_{s\beta} 
  &=& -\rmi \mu_0 \Omega_{s\beta} \bfH_{s\beta} 
  \\
  \rot \bfH_{s\beta} 
  &=& \rmi \epsilon_0 \Omega_{s\beta} \bfE_{s\beta}
\end{eqnarray}
where the $\Omega_{s\beta}$ are the real eigenvalues, $\mu_0$ and
$\epsilon_0$ the permeability and permittivity of vacuum.  For a given
$\beta \in [ 0, 2\pi/d [$, the set of eigen electric fields $\bfE_{s\beta}$
($s \in \mathbb{N}$) is an orthogonal basis 
of the divergence-free subset of $H(\rot, \cV_0)$ 
(this is the Hilbert space of square-integrable fields on $\cV_0$ 
which have a square-integrable curl).
Any divergence-free field $\bfE_\beta(\bfr,t)$ on $\cV_0$ can be
decomposed on the $\bfE_{s\beta}$ with  coefficients $V_{s\beta}$, 
\begin{equation}
  \bfE_\beta(\bfr,t)=\sum_{s \in \NN} V_{s\beta}(t) \bfE_{s\beta}(\bfr).
  \label{eq:decomposition-1}
\end{equation}
This decomposition can be continued to $\cV_\ZZ$ thanks to the
Floquet condition, and the continuation field satisfies the Floquet
condition as well (possibly with a Gibbs phenomenon, 
depending on the field regularity).

An arbitrary function $G$ in the propagating structure does not, in
general, satisfy the Floquet condition. But it can be expanded on
functions $G_\beta$ satisfying the Floquet condition. Indeed, the
Fourier series on the variable $\beta$, with $x$ as a parameter,
\begin{equation}
G_\beta(x,t):=\sum_{n \in \ZZ}G(x+nd,t)\rme^{\rmi n \beta d}
\label{eq:beta-transform}
\end{equation}
satisfies the Floquet condition with respect to $x$. The $n$-th Fourier
coefficient in the series is given by
\begin{equation}
G(x+nd,t)=\frac{1}{2\pi}\int_{\beta d=0}^{2\pi}G_\beta(x,t)\rme^{-\rmi
  n\beta d}\rmd (\beta d),
\label{eq:inv-beta-transform}
\end{equation}
which for $n=0$ yields exactly the requested expansion. The transform
(\ref{eq:beta-transform}), hereafter referred to as the
$\beta$-transform \cite{Gelfand50,Kuznetsov80,Ryskin09,Zayed11}, 
decomposes any function
into a superposition of functions satisfying the Floquet condition. It
will be central in what follows. Relation
(\ref{eq:inv-beta-transform}) is the inverse $\beta$-transform.

Using eq.~(\ref{eq:decomposition-1}), we finally obtain
\begin{equation}
\bfE(\bfr,t)=\frac{1}{2\pi}\sum_{s \in \mathbb{N}}\int_{\beta
  d=0}^{2\pi}V_{s\beta}(t) \bfE_{s\beta}(\bfr)\rmd (\beta d)   .
\label{eq:decomposition-2}
\end{equation}

This decomposition holds for any free electric field in the structure,
e.g.\  propagating modes, evanescent modes or  superposition of
these. The magnetic field expansion is written
\begin{equation}
\bfH_\beta(\bfr,t)=\rmi\sum_{s \in \mathbb{N}}I_{s\beta}(t) \bfH_{s\beta}(\bfr)
\label{eq:decomposition-H}
\end{equation}
with the factor $\rmi$ for later convenience.

Our initial choice to consider the propagating field independently
holds only if it is decoupled from the space-charge field. This latter
adds the curl-free term $-\nabla \phi$ to the expression of $\bfE$
where $\phi$ satisfies the Poisson equation $\Delta \phi = -
\rho/\epsilon_0$. The $\beta$-transform commutes both with time and
space derivatives, so Maxwell equations are valid for $\bfE_\beta$ and
$\bfH_\beta$ as well. Using the $\beta$-transform of $\phi$, the
general form of the electric field in the presence of charged
particles is
\begin{equation}
  \bfE_\beta(\bfr,t)=\sum_{s \in \mathbb{N}}V_{s\beta}(t) \bfE_{s\beta}(\bfr) -
  \nabla \phi_\beta.
\label{eq:decomposition-E}
\end{equation}
Using Maxwell equations, the field decompositions
(\ref{eq:decomposition-E}) and (\ref{eq:decomposition-H}), and the
definition of eigenfields $\bfE_{s\beta}$ and $\bfH_{s\beta}$, we
obtain
\begin{eqnarray}
  - \sum_s I_{s\beta} \Omega_{s\beta} \bfE_{s\beta} 
  &=& 
  \sum_s \dot{V}_{s\beta}
    \bfE_{s\beta} + \frac{\bfJ_\beta}{\epsilon_0} 
    - \frac{\partial \nabla \phi_\beta}{\partial t} , 
    \\
  \sum_s V_{s\beta} \Omega_{s\beta} \bfH_{s\beta} 
  &=& 
  \sum_s \dot{I}_{s\beta} \bfH_{s\beta} ,
\end{eqnarray}
where $\bfJ$ is the current density.  We multiply the first expression
by the conjugate of the field $\bfE^*_{s'\beta}$ and the second
expression by $\bfH^*_{s'\beta}$. Integrating the resulting
expressions over the volume $\cV_0$ yields vanishing terms for $s \ne
s' $. Otherwise we exhibit the electric $\frac{1}{2}\epsilon_0
\int_{\cV_0} |\bfE_{s\beta}|^2 \rmd^3 \bfr$ and magnetic
$\frac{1}{2}\mu_0\int_{\cV_0} |\bfH_{s\beta}|^2 \rmd^3 \bfr$ energies
stored in one period for the corresponding propagation mode. Both
energies are equal to half the total stored energy $N_{s\beta}$ of the
basis function. The volume integral involving $\nabla \phi_\beta$ is
transformed to a surface integral thanks to the identity $\nabla \cdot
(\phi_\beta \bfE_{s'\beta}^*) = (\nabla \phi_\beta) \cdot
\bfE_{s'\beta}^* + \phi_\beta \nabla \cdot \bfE_{s'\beta}^*$. The
electric field basis functions are divergence-free, so only the
surface integral $\int_{\partial \cV_0}\phi_\beta \bfE_{s'\beta}^*
\cdot \bfe_\perp \rmd S$ remains.  The surface integral over the two
cross-sections of the waveguide vanishes because both $\phi_\beta$ and
$\bfE_{s'\beta}$ satisfy the Floquet condition. The lateral metallic
parts are at imposed time-invariant potentials, a property
$\phi_\beta$ inherits, so the time derivative of the corresponding
surface integral vanishes. This concludes our demonstration that the
evolution of the $V_{s\beta}$ and $I_{s\beta}$ representing the
propagating field is decoupled from the space-charge field
\begin{eqnarray}
  \dot{V}_{s\beta} + \Omega_{s\beta} I_{s\beta} 
  & = &
  - \frac{1}{N_{s\beta}}\int_{\cV_0} \bfJ_\beta \cdot \bfE^*_{s\beta} \rmd^3 \bfr, 
  \label{eq:couplage}
  \\
  \dot{I}_{s\beta}-\Omega_{s\beta} V_{s\beta} 
  &= & 
  0.
\label{eq:evolution-Ibeta}
\end{eqnarray}
We perform the inverse $\beta$-transform of (\ref{eq:couplage}) and
(\ref{eq:evolution-Ibeta}). Like the usual Fourier coefficients, the
$\beta$-transform of a product is the convolution of its factor
transforms.  The source term in Maxwell-Amp\`ere is transformed by
(\ref{eq:beta-transform}) into an integral over the complete volume
$\cV_\ZZ$~:
\begin{eqnarray}
   \int_{\cV_0} \bfJ_\beta \cdot \bfE^*_{s\beta}\rmd^3 \bfr
    = 
   \int_{\cV_0} \sum_n
      \bfJ(\bfr + n d \bfe_x) \cdot \bfE^*_{s\beta}(\bfr) \rme^{\rmi n\beta d} \rmd^3 \bfr 
  \nonumber \\
   =   
  \int_{\cV_0} \sum_n 
     \bfJ(\bfr + n d \bfe_x) \cdot \bfE^*_{s\beta}(\bfr + n d \bfe_x) \rmd^3 \bfr
  \nonumber \\
   =  
  \int_{\cV_\ZZ} \bfJ \cdot \bfE^*_{s\beta}\rmd^3 \bfr , \nonumber
\end{eqnarray}
where the second equality follows from the Floquet property of $\bfE_{s\beta}$.
Finally, using the inverse $\beta$-transform, 
\begin{eqnarray}
  \dot{V}_{sn} + \sum_m \Omega_{sm} I_{s, n-m} 
  & = & 
  - \int_{\cV_\ZZ} \bfJ \cdot \bfF_{s, -n} \rmd^3 \bfr ,
  \label{eq:ev-temp-V}
  \\
  \dot{I}_{sn} - \sum_m \Omega_{sm} V_{s, n-m} 
  & = & 
  0 ,
  \label{eq:ev-temp-I}
\end{eqnarray}
where
\begin{equation}
  \bfF_{s, -n}:=\frac{1}{2\pi} \int^{2\pi}_0
  \frac{\bfE^*_{s\beta}}{N_{s\beta}}\rme^{\rmi n\beta d}\rmd (\beta d)
\label{eq:definition-Fsn}
\end{equation}
whose $\beta$-transform is $\bfF_{s\beta} = \bfE_{s\beta}/N_{s\beta}$.

The electric and magnetic fields, given by
\begin{eqnarray}
  \bfE(\bfr,t) & = & \sum_{s, n} V_{sn}(t) \bfE_{s, -n}(\bfr) - \nabla \phi(\bfr,t), \\
  \bfH(\bfr,t) & = & \rmi\sum_{s, n} I_{sn}(t) \bfH_{s, -n}(\bfr),
\end{eqnarray}
are rewritten
\begin{eqnarray}
  \bfE(\bfr,t) & = & \sum_{s, n} V_{sn}(t) \bfE_{s0}(\bfr - n d \bfe_x) - \nabla \phi(\bfr,t), \\
  \bfH(\bfr,t) & = & \rmi\sum_{s, n} I_{sn}(t) \bfH_{s0}(\bfr - n d \bfe_x).
\end{eqnarray}

Each eigenfunction $\bfH_{s\beta}$ derives from a vector potential
$\bfA_{s\beta}$
\begin{equation}
  \mu_0 \bfH_{s\beta}=\rot \bfA_{s\beta} .
\end{equation}
The inverse $\beta$-transform yields $\mu_0 \bfH_{sn}=\rot \bfA_{sn}$. Using
(\ref{eq:decomposition-H}), we find
\begin{equation}
\bfA = \rmi\sum_{s, n} I_{sn} \bfA_{s, -n} + \nabla u
\label{eq:decomposition-A}
\end{equation}
where $u$ is any function of space corresponding to a gauge choice.

These expressions offer an interpretation for the index $n$. The
electric field at a particular position is the superposition of the
field shape $\bfE_{s0}$ generated by the adjacent cells of the
periodic structure with amplitudes $V_{sn}$.  Therefore $V_{sn}$
(resp.\  $I_{sn}$) can be seen as the contribution of the $s$-th
eigenmode in cell $n$ to the electric (resp.\  magnetic) field. The
propagation is described by linearly coupled harmonic
oscillators. Cell $n$ is coupled with cell $m$ through the coefficient
$\Omega_{s, n-m} = \frac{1}{2\pi} \int_{-\pi}^{\pi} \Omega_{s\beta}
\rme^{\rmi (n-m) \beta d} \rmd(\beta d)$, viz.\  the corresponding
Fourier coefficient of the dispersion curve $\Omega_{s\beta}$. The
reciprocity condition $\Omega_{s, -\beta} = \Omega_{s\beta}$ ensures
$\Omega_{s, -n} = \Omega_{sn} \in \RR^+$. The same condition on
the fields reads $\bfE_{s, -\beta}=\bfE^*_{s\beta}$, therefore
$\bfE_{sn}
  = \frac{1}{2\pi} \int_{-\pi}^{+\pi}\bfE_{s\beta}^*\rme^{\rmi n\beta d} \rmd (\beta d)
  = \frac{1}{\pi} \Re \left( \int_{0}^{+\pi} \bfE_{s\beta}^* 
             \rme^{\rmi n\beta d} \rmd (\beta d) \right) $ 
  is a vector with real components only. 
Symmetrically, $\rmi \bfH_{sn}$ has only real components. 
In summary, $V_{sn}$, $I_{sn}$,
$\bfE_{sn}$, $\bfF_{sn}$, $\rmi\bfH_{sn}$, $\rmi \bfA_{sn}$,
$\Omega_{sn}$, $N_{sn}$ are all real quantities.

Starting from this model, we can now construct a Hamiltonian for the
electromagnetic field. The standard \cite{Hartemann02} Hamiltonian
$\frac{1}{2}\int_\cV (\epsilon_0 |\bfE|^2 + \mu_0 |\bfH|^2 ) \rmd^3
\bfr$ results from the Lorentz force and the Maxwell-Amp\`ere
equation. In this work, we want to make no assumption on the final
form of the electromagnetic Hamiltonian because corrective terms might
be needed. Consequently, we cannot use it as a starting
point. Instead, the previous development provides a Hamiltonian for
the fields alone, independently from how they are coupled with the
electrons. The vectors $V_s = (\dots V_{sn} \dots )$ and $I_s = (\dots
I_{sn} \dots )$ represent the state of the fields, and their time
evolutions derive from the Hamiltonian (with conjugate variables $V,
I$)
\begin{equation}
  \mathcal{H}_\mathrm{em}(V_s,I_s)_{s \in \mathbb{N}}
  = \frac{1}{2} \sum_s  ( V_sQ_sV_s^t+I_sQ_sI_s^t  )  ,
\label{eq:Hem}
\end{equation}
where $\cdot^t$ denotes vector transpose, and $Q_s$ is the infinite
matrix with entries $(Q_s)_{n,m} = \Omega_{s, n-m}$ for $(n,m) \in
\mathbb{Z}^2$. Till now, the basis functions $\bfE_{s\beta}$ and
$\bfH_{s\beta}$ can have arbitrary physical dimensions. Adopting
(\ref{eq:Hem}) as a Hamiltonian, however, results in $V_s Q_s V_s^t$
and $I_s Q_s I_s^t$ being energies. Hence $V_s$ and $I_s$ have the
same dimensions.  Eqs. (\ref{eq:ev-temp-V})-(\ref{eq:ev-temp-I})
impose $\Omega_{sn}$ to be a frequency. Therefore $V_s$ and $I_s$ have
the dimension of the square root of an action (energy divided by
frequency). The electric and magnetic field have their standard
dimensions and $N_{s\beta}$ has the dimension of a frequency.

The Hamiltonian of the electron in the  field of a given 4-potential 
$(\phi,\bfA)$ is \cite{LandauLifshitzVol2}
\begin{equation}
  \mathcal{H}_\mathrm{el}(\bfp,\bfr)
  =
  \sqrt{m^2c^4+c^2 | \bfp - e\bfA(\bfr) |^2} + e\phi(\bfr)
\label{eq:hamiltonien-el}
\end{equation}
where $e$ and $m$ are the electron charge and mass, $c$ is the
celerity of light, and the momentum $\bfp$ is the canonical
conjugate of the electron position $\bfr$. 
The complete Hamiltonian for the fields and $M$ electrons is
necessarily the sum of the purely electromagnetic
Hamiltonian $\mathcal{H}_\mathrm{em}$ with the Lorentz force dynamic Hamiltonian
$\mathcal{H}_\mathrm{el}$, where $\bfA$ is a function of the $I_{sn}$ as given
by (\ref{eq:decomposition-A}) 
and $\phi$ generates the space-charge field,
\begin{equation}
  \mathcal{H}(\bfp_j,\bfr_j, V_s,I_s) _{\lbrace 1 \leq j \leq M, s \in \NN \rbrace }
  = 
  \sum_j \mathcal{H}_\mathrm{el}(\bfr_j,\bfp_j{,I_s}) + \mathcal{H}_\mathrm{em}.
\label{eq:hamiltonien}
\end{equation}
The complete physics of the system is governed by it, including the
source term in the Maxwell-Amp\`ere equation. This source term has not
been used to build the Hamiltonian and is equivalent to eq.~(\ref{eq:ev-temp-V}). 
In the presence of a single electron with
position $\bfr(t)$ and velocity $\bfv(t)$, 
we have \cite[\S 29]{LandauLifshitzVol2} $\bfJ(\bfr,t) = e \bfv(t) \, \delta (
\bfr-\bfr(t) ) $, and (\ref{eq:ev-temp-V}) simplifies into
\begin{equation}
  \dot{V}_{sn} =
  - e \bfv \cdot \bfF_{s, -n}(\bfr)
  - \sum_m \Omega_{sm} I_{s, n-m}.
\label{eq:evol-V-Max}
\end{equation}
Its derivation from $\mathcal{H}$ is obtained  considering that
$V_{sn}$ is the generalized momentum canonically conjugate 
to the generalized coordinate $I_{sn}$,
\begin{eqnarray}
  \dot{V}_{sn}
  & = & 
  - \derivep{\mathcal{H}}{I_{sn}} 
  \nonumber \\
  & = & 
  - \derivep{\mathcal{H}_\mathrm{em}}{I_{sn}} 
  + \frac{\bfp - e \bfA}{\sqrt{m^2 c^4 + c^2 | \bfp - e \bfA |^2}} \cdot e \rmi \bfA_{s, -n} 
  \nonumber \\
  & = & e \bfv \cdot \rmi \bfA_{s, -n}(\bfr)  - \sum_m \Omega_{sm} I_{s, n-m}
\label{eq:evol-V-ham}
\end{eqnarray}
Here we used the relation $\bfv = \dot{\bfr} = \partial
\mathcal{H}_\mathrm{el} / \partial \bfp 
= c^2 (\bfp - e \bfA)/ \sqrt{m^2 c^4 + c^2 | \bfp - e \bfA |^2}$, 
  that derives directly from the electron
Hamiltonian (\ref{eq:hamiltonien-el}).

Eqs. (\ref{eq:evol-V-Max}) and (\ref{eq:evol-V-ham}) are
incompatible unless 
\begin{equation}
\bfF_{s n}(\bfr)=-\rmi \bfA_{sn}(\bfr).
\end{equation} 
The $\beta$-transform gives
$\bfE_{s\beta}/N_{s\beta} = \bfF_{s\beta} = -\rmi\bfA_{s\beta}$. By
definition, 
$\bfE_{s\beta} = - \rmi \Omega_{s\beta} \bfA_{s\beta} - \nabla u_{s\beta}$, 
so the two equations reconcile when
\begin{equation}
  N_{s\beta} = \Omega_{s\beta} 
  \quad \text{and} \quad
  \nabla u_{s\beta} = 0.
\end{equation}
It is always possible to choose basis functions satisfying the Coulomb
gauge, which is the second condition. The first condition clarifies
the question of the electromagnetic energy. In this work, it is given
by the Hamiltonian $\mathcal{H}_\mathrm{em}$ while the usual Poynting
energy is $\frac{1}{2}\int_{\cV_\ZZ} ( \epsilon_0 |\bfE|^2 + \mu_0
|\bfH|^2 ) \rmd^3 \bfr$.  In the latter expression, consider the
electric part $\mathcal{E} = \frac{1}{2}\epsilon_0 \int_{\cV_\ZZ}
|\bfE|^2 \rmd^3 \bfr = \frac{1}{2}\epsilon_0 \sum_n \int_{\cV_0}
|\bfE(\bfr+nd\bfe_x)|^2 \rmd^3 \bfr$. Using Parseval's relation, this
is also $\frac{1}{2} \epsilon_0 \int_{\cV_0} \int_0^{2\pi}
|\bfE_\beta|^2 \rmd (\beta d) \rmd^3 \bfr$. Now, using the
decomposition (\ref{eq:decomposition-E}) of the electric field, we
find $\mathcal{E} = \frac{1}{2}\epsilon_0 \int_{\cV_0} \int_0^{2\pi} |
\sum_s V_{s\beta}\bfE_{s\beta} |^2 \rmd (\beta d) \rmd^3 \bfr$.
Orthogonality of the eigenmodes reduces this expression to
$\mathcal{E} = \frac{1}{2} \sum_s \int_0^{2\pi} N_{s\beta}
|V_{s\beta}|^2 \rmd (\beta d)$.  A similar calculation applies to the
magnetic energy.  Finally, using the inverse $\beta$-transform and the
condition $N_{s\beta} = \Omega_{s\beta}$, we find that both
expressions for the energy, $\mathcal{H}_\mathrm{em}$ and Poynting's,
are equal:
\begin{equation}
  \frac{1}{2} \int_{\cV_\ZZ} 
    ( \epsilon_0 |\bfE|^2 + \mu_0 |\bfH|^2 ) \rmd^3 \bfr =
  \frac{1}{2} \sum_s ( V_s Q_s V_s^t + I_s Q_s I_s^t ).
\end{equation}

In this work, we constructed a Hamiltonian describing the electrons
coupled with the electromagnetic fields propagating in a periodic
structure. This coupled oscillator model may be viewed as a
generalization of the telegraph delay line where each period consists
of one inductor and capacitor, coupled with their nearest neighbors
and electrons \cite{Pierce55}. In contrast, in the present work all
periods are coupled with each other. With this model, we can compute
the time-dependent behavior of a helix traveling-wave tube, much
faster than industrial PIC codes
\cite{Bernardi_IEEE11b}. {The reason is that the number of
  degrees of freedom to describe the propagating wave with sufficient
  accuracy in finite-difference time-domain 
  or finite element techniques amounts to tens of
  thousands per pitch \cite{AndreAissi10}. In contrast, the beam is
  interacting with only one mode in a traveling wave tube, so that the number of
  degrees of fredom in the present theory is reduced to 
  only one conjugate pair per pitch, 
  the $(V_n, I_n)$ of the contemplated mode.}

While we focused on divergence-free, propagating fields, the space
charge field may be taken into account by adding the ``space charge
Hamiltonian'' of two or more interacting electrons or by adding
Darwin's first-order relativistic approximation \cite[\S
12.6]{Jackson99} (Green functions are needed to take into account the
waveguide boundary condition). In this action-at-a-distance approach
with two electrons interacting through their electrostatic field, the
problem of infinite electromagnetic mass does not occur. This
Hamiltonian is formally independent of time but depends explicitly on
space through the $\bfA_{sn}(\bfr)$. Indeed the boundary conditions --
for example a corrugated metallic wall -- enable the field to exchange
momentum but not energy. Maxwell's equations and the Lorentz force
correctly account for the recoil from the radiated field.

{Self-acceleration can exist in the framework of a
  Hamiltonian description when it is the sum of two terms with
  opposite signs. In this case, the negative term of the system can
  indefinitely yield energy to the positive term while satisfying
  energy conservation.}  On the contrary, our Hamiltonian
{(\ref{eq:hamiltonien})} is the sum of two unconditionally
positive contributions which are minimum when the system is at rest, a
situation precluding self-acceleration. As a consequence of the
Hamiltonian, the source term in the equation of Maxwell-Amp\`ere and
the Lorentz force consistently express how the electron is coupled
with the propagating field.

PB was supported by a CIFRE stipend from the French 
Minist\`ere de l'Enseignement Sup\'erieur et de la Recherche.
NR enjoyed Thales' hospitality.
{A referee is warmly thanked for her/his recommendations.}

\end{document}